\documentclass[aps,prb,twocolumn,floatfix] {revtex4-2}
\usepackage{graphicx,amsfonts}
\usepackage{bm,color}
\usepackage{multirow}
\usepackage{amssymb,amsmath,hyperref}
\usepackage{wrapfig}
\usepackage{float}

\begin{document}

\title{Microwave quantum diode}

\author{Rishabh Upadhyay$^1$}
\affiliation{Pico group, QTF Centre of Excellence, Department of Applied Physics, Aalto University School of Science, P.O. Box 13500, 00076 Aalto, Finland}
\author{Dmitry S. Golubev$^1$}
\affiliation{Pico group, QTF Centre of Excellence, Department of Applied Physics, Aalto University School of Science, P.O. Box 13500, 00076 Aalto, Finland}
\author{Yu-Cheng Chang$^1$}
\affiliation{Pico group, QTF Centre of Excellence, Department of Applied Physics, Aalto University School of Science, P.O. Box 13500, 00076 Aalto, Finland}
\author{George Thomas $^1$}
\affiliation{Pico group, QTF Centre of Excellence, Department of Applied Physics, Aalto University School of Science, P.O. Box 13500, 00076 Aalto, Finland}
\author{Andrew Guthrie$^1$} 
\affiliation{Pico group, QTF Centre of Excellence, Department of Applied Physics, Aalto University School of Science, P.O. Box 13500, 00076 Aalto, Finland}
\author{Joonas T. Peltonen$^1$}
\affiliation{Pico group, QTF Centre of Excellence, Department of Applied Physics, Aalto University School of Science, P.O. Box 13500, 00076 Aalto, Finland}
\author{Jukka P. Pekola$^1$}
\affiliation{Pico group, QTF Centre of Excellence, Department of Applied Physics, Aalto University School of Science, P.O. Box 13500, 00076 Aalto, Finland}

\begin{abstract}

The fragile nature of quantum circuits is a major bottleneck to scalable quantum applications. Operating at cryogenic temperatures, quantum circuits are highly vulnerable to amplifier backaction and external noise. Non-reciprocal microwave devices such as circulators and isolators are used for this purpose. These devices have a considerable footprint in cryostats, limiting the scalability of quantum circuits. We present a compact microwave diode architecture, which exploits the non-linearity of a superconducting flux qubit. At the qubit degeneracy point we experimentally demonstrate a significant difference between the power levels transmitted in opposite directions. The observations align with the proposed theoretical model. At -99~dBm input power, and near the qubit-resonator avoided crossing region, we report the transmission rectification ratio exceeding $90\%$ 
for a $50$ MHz wide frequency range from $6.81$ GHz to $6.86$ GHz, and over $60\%$ for the $250$~MHz range from $6.67$ GHz to $6.91$ GHz.  The presented architecture is compact, and easily scalable towards multiple readout channels, potentially opening up diverse opportunities in quantum information, microwave read-out and optomechanics.

\end{abstract}

\maketitle

\textbf{Introduction}:

Quantum engineering, a dynamic discipline bridging the fundamentals of quantum mechanics and established engineering fields 
has developed significantly in the past few decades. 
Two-level systems such as superconducting quantum bits are the building blocks of quantum circuits. 
Qubits of this type are the most researched and promising candidates for the realization of quantum information processing \cite{JohnClarke,G.Wendin,M.H.Devoret,XiuGu,Ronzani2019}. 
The characteristics of the superconducting qubits such as eigen energies, non-linearity, coupling strengths etc. can be tailored more easily by adjusting the design parameters \cite{S.E. Rasmussen, T.P.Orlando} than some other two-level microscopic quantum systems \cite{Daniel Loss, B. E. Kane, Atsushi Goto, J. R. Petta, Z.Wang, R. Hanson}. Qubits are engineered to have large non-linearity, which makes it possible to selectively address and control them \cite {JohnClarke, M.H.Devoret, T.P.Orlando, J.E.Mooij}. This dynamic property makes superconducting qubits a strong candidate for plethora of applications.

Quantum devices operate at low temperatures and require good isolation from external noises.
Microwave devices, such as circulators and isolators, protect quantum circuits by unidirectionally routing the output signal, whilst simultaneously isolating noise from the output channel back to the quantum circuit. Their non-reciprocal character relies on the properties of ferrites \cite{D.M.Pozar, A.R. Hamann, Toshiro Kodera}. 
Ferrite-based non-reciprocal devices are bulky \cite{D.M.Pozar, A.R. Hamann, Toshiro Kodera}, and they cannot 
be positioned near the quantum circuit because they require strong magnetic fields. 
This limits the scalability of cryogenic quantum circuits \cite{D.M.Pozar, A.R. Hamann, Giovanni Viola, Leonardo Ranzani}. 
Various ferrite-free alternatives have been proposed in literature. 
These proposals rely on the properties of noble materials \cite{Giovanni Viola}, non-linear behavior of artificial atoms  \cite{A.R. Hamann}, 
dc superconducting quantum interference devices (dc-SQUID) \cite{Archana Kamal}, and arrays of Josephson junctions (JJ's) \cite{Leonardo Ranzani,Yaakobi, Lecocq, Brien}. Our device is a proof of concept (PoC), potentially useful in the applications relevant to microwave read-out components in the field of superconducting quantum circuits. 

\begin{figure*}[ht]
\includegraphics [width = 0.8\textwidth] {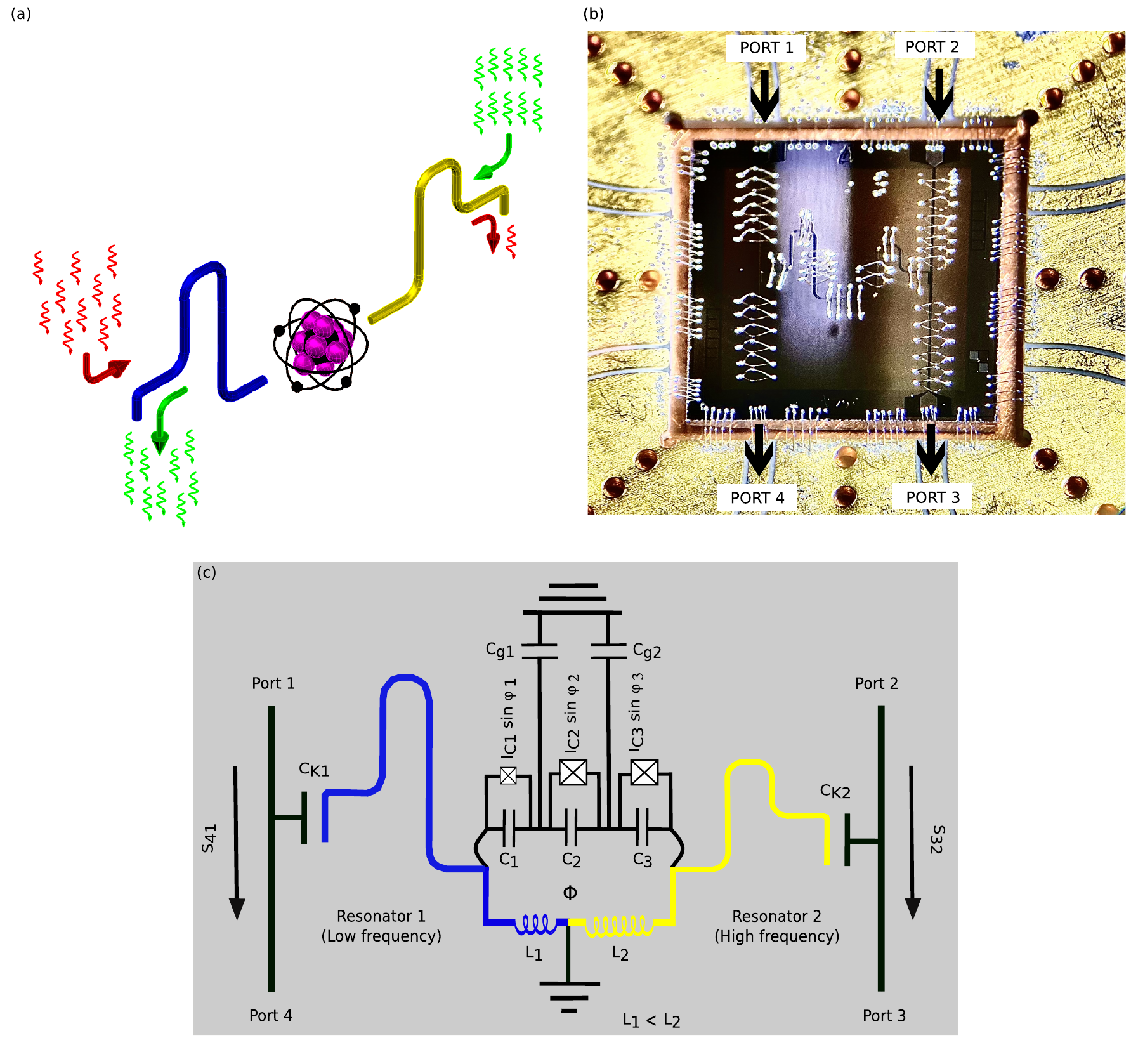}
\caption{(a) Conceptual representation of the device as an artificial atom coupled to two resonators. 
The photons pass easily from the right side (green arrows) to the left side, whereas those coming from left side (red arrow) are mainly reflected back.  
(b) Optical microscope image of the device bonded on gold plated copper sample stage. 
The RF signal enters either via port 1 or port 2. Port 3 and Port 4 are the output ports. 
The black arrows indicate the direction of signal propagation. 
(c) Circuit model of the device. Here, $\Phi_{\rm }$ is the external magnetic flux threading through the qubit loop.
}
\label{Device_concept}
\end{figure*}

In this work, we present a robust and simple on-chip microwave diode demonstrating transmission rectification based on a superconducting flux qubit \cite {J.E.Mooij}. 
The concept of the device is shown in Fig.~\ref{Device_concept} (a). The flux qubit is inductively coupled to two superconducting resonators 
of different lengths with different coupling strengths. 
The design details are reported in Section ~\ref{Design and measurement scheme}. 
Probing the qubit at the half-flux (degeneracy point) with one tone-spectroscopy, 
we observe identical patterns of transmission coefficient for signals propagating in the opposite directions, which are shifted by 5 dB in power. This power induces the jump from the low energy to the high energy state of the qubit. This shift indicates the non-reciprocal behaviour in our device, expressed in terms of transmission rectification ratio (R) in this article. We study the transmission rectification ratio, $R$, under different injected microwave powers and in a wide range of frequencies and magnetic fluxes applied to the qubit. 
Due to its strong non-reciprocity, our device could potentially be utilized as a ferrite free on-chip isolator in a microwave readout scheme \cite{Archana Kamal, Leonardo Ranzani, Giovanni Viola,  Brien}.  The strong non-reciprocal behaviour observed in the reported device is relevant in the field of circuit quantum thermodynamics (c-QTD) \cite{ Sai, Kosloff} to facilitate and manage the heat flow in superconducting quantum circuits \cite{Erdman, Menczel, Giazotto, Miranda, Robnagel, B.Karimi, Jorden2020, Robnagel, LuY}. Moreover, its compact size makes it suitable for multiple read-out channels. The possibility to control the transmission rectification ratio with tiny magnetic fields provides an additional advantage.

\begin{figure*}[ht]
\includegraphics [width = 0.85\textwidth] {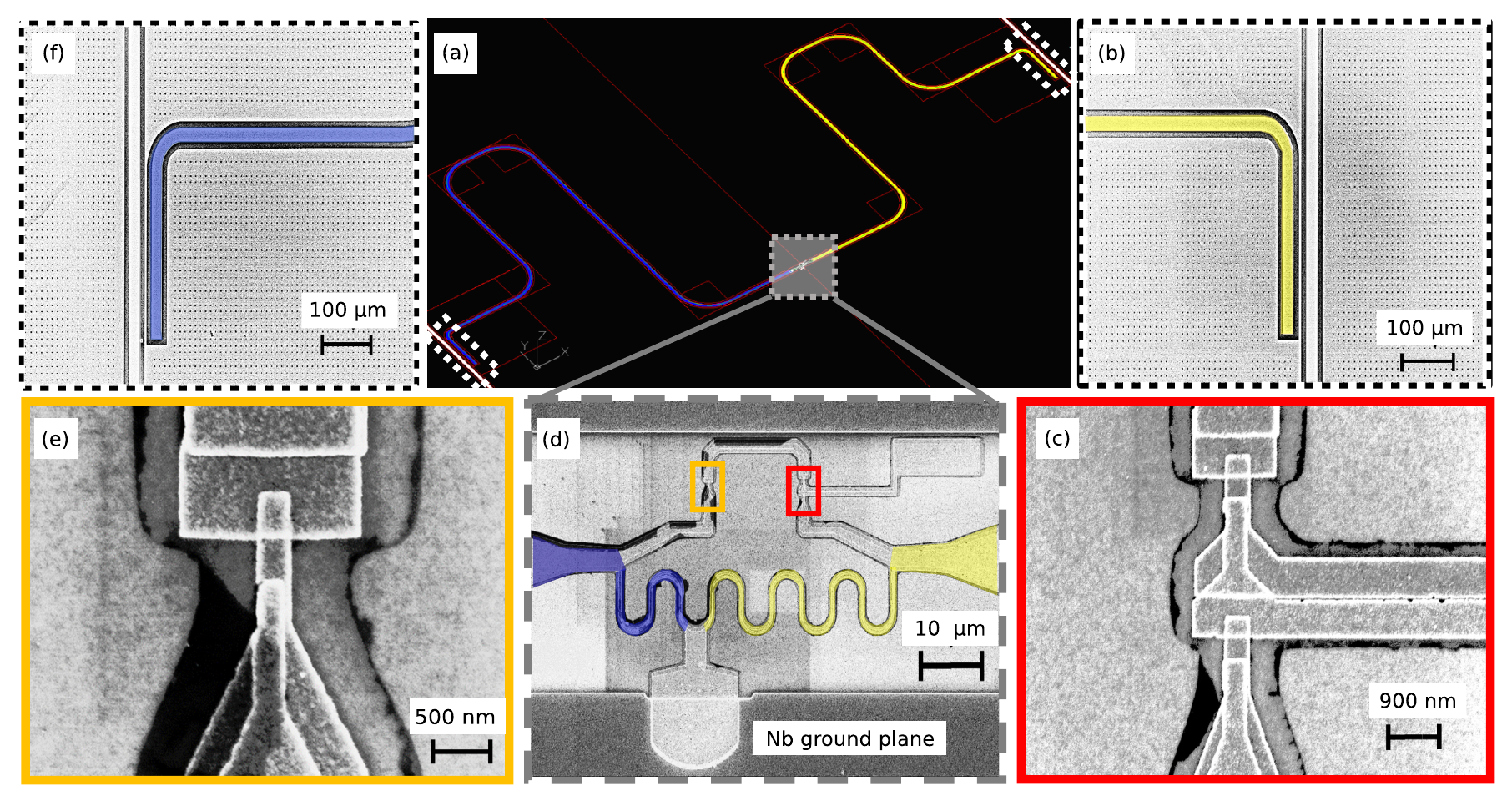}
\caption{(a) The layout of the device exhibiting the two resonators and the three-junction superconducting flux qubit at the center. 
The right and the left resonators are shown by yellow and by dark-blue colors respectively. 
The areas enclosed with white dotted lines at the top-right and bottom-left corner show the locations of the capacitors coupling the right 
and the left resonators with the feedlines. In panels (b) and (f) we show the magnified images of these capacitors.
The gray shaded area close to the center in panel (a) shows a three-terminal flux qubit. 
Its zoomed image is shown in panel (d). The qubit is coupled to both resonators via the local inductances to the left and right, highlighted with different colors. 
(c) An enlarged electron micrograph of the two big junctions of the flux qubit. 
(e) Electron microscope image of the smaller qubit junction. 
}
\label{Device_design}
\end{figure*}

\begin{figure*}[ht]
\includegraphics  [width = 0.9\textwidth] {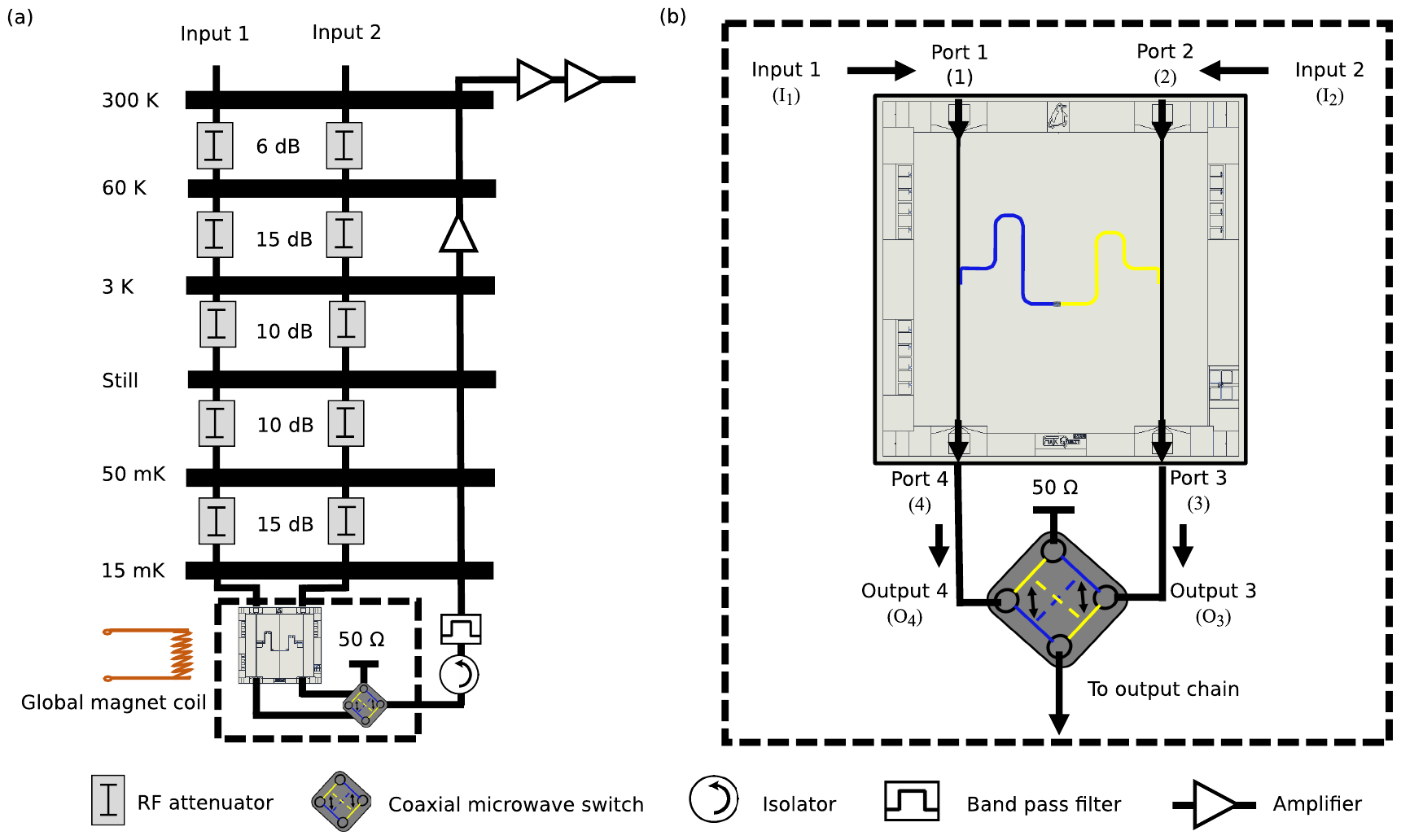}
\caption{Measurement setup.
(a) Different temperature stages of the fridge with respective attenuation at each stage. 
The attenuation in the input lines 1 and 2 are nominally identical. 
(b) An enlarged image of the sample setup at the mixing chamber. 
Input 1 connects to the port 1 and input 2 --- to the port 2. 
The output ports 3 and 4 are connected to the low-temperature coaxial microwave switch. 
The switch is driven by DC (V) bias. 
It connects one of its input ports (input 4 in the figure) to the output and, at the same time, terminates the other input (input 3 in the figure) at 50 $\mathrm{\Omega}$.}
\label{Measurement_setup}
\end{figure*}

\section{Design and measurement scheme}
\label{Design and measurement scheme}

{\it Design}.
An optical microscope image of the device is shown in Fig.~\ref{Device_concept} (b),
and the circuit diagram in Fig.~\ref{Device_concept} (c). 
The core element of the device is a three-junction superconducting flux qubit \cite {J.E.Mooij, Yan}.
Its magnified image is presented in Fig.~\ref{Device_design} (d). 
The superconducting loop of the flux qubit contains Josephson junctions. The areas of junctions numbered as 2 and 3 are nominally the same, whereas the junction 1 is smaller. 
Hence, the critical currents of junctions 2 and 3 are nominally equal, $I_{C2}=I_{C3}=I_{C}$, 
and that of the first junction is given by $I_{C1}=\alpha I_{C}$, where we estimate $\alpha=0.632$. 
The flux qubit has two superconducting islands, which we number as the island 1 and the island 2. 
The total capacitance of island 1 is given by $C_{G1}= C_1+C_2+C_{g1}$, where $C_1$, $C_2$ are the capacitances of the junctions 1, 2 
and $C_{g1}$ the capacitance to the ground plane. Analogously, the total capacitance of the second island is $C_{G2}= C_2+C_3+C_{g2}$. 
The flux qubit is inductively coupled to the resonators 1 and 2 via the inductances $L_1$ and $L_2$, 
realized as a superconducting aluminum wire divided in two parts by the grounding electrode, see Fig. ~\ref{Device_design} (d). 
The corresponding coupling constants are proportional to the inductances, $g_j\propto L_j$ ($j=1,2$)  
\cite{Abdumalikov, J.Bourassa, R.Upadhyay, T.Miyanaga}. 
The left resonator (resonator 1) having the designed frequency  
$f_1=6.5$~GHz, is more weakly coupled to the qubit, while
the right resonator (resonator 2), with a designed frequency of $f_2=7.5$~GHz, 
has stronger coupling, i.e. $g_1<g_2$. The frequencies $f_1$ and $f_2$ cited above are the nominal values
expected in the limit of $\lambda/4$-resonators, which is achieved at $L_1,L_2\to 0$.
At finite qubit resonator coupling they may shift downwards by an unknown value. 
Both resonators are coupled to the transmission lines via nominally equal coupling capacitances as shown in Fig.~\ref{Device_design} (b),(f). 
Further details of device and its fabrication are reported in Section ~\ref{Appendix_A}.

{\it Measurement scheme}.
The device is wire-bonded to a gold-coated printed circuit board as shown in Fig.~\ref{Device_concept} (b). 
The RF spectroscopy measurements have been performed in a cryo-free dilution refrigerator at the base temperature of 15~mK. 
The measurement setup is schematically shown in Fig.~\ref{Measurement_setup} (a). 
We have used two input lines with identical attenuation at different temperature stages of the fridge. 
At the output end, we have used a commercial coaxial microwave switch, which allowed us to connect either port 3 or port 4 of the device to a single output channel. 
Under applied DC bias this coaxial microwave switch connects the output channel to the desired port and terminates the other port to $50~\Omega$ to avoid reflections, as shown in Fig.~\ref{Measurement_setup} (b). 
This setup eliminates unwanted differences in microwave transmission through output ports 3 and 4, which might otherwise affect the measurements.  
The signal then follows an output line that includes two isolators embedded at the mixing chamber. At the 4~K stage the signal is amplified by a low noise HEMT amplifier by 42~dB. Furthermore, the signal is amplified at room temperature by 52~dB. 
For the flux bias of the qubit we have used a DC-driven global magnet coil at the mixing chamber, as shown in Fig.~\ref{Measurement_setup} (a). 
We have characterized the device using one tone and two tone-spectroscopy methods, briefly discussed in Section ~\ref{Experimental observations} and in Appendix~\ref{Two tone-Spectroscopy}.

\begin{figure}[ht]
\includegraphics [width = 0.9\columnwidth] {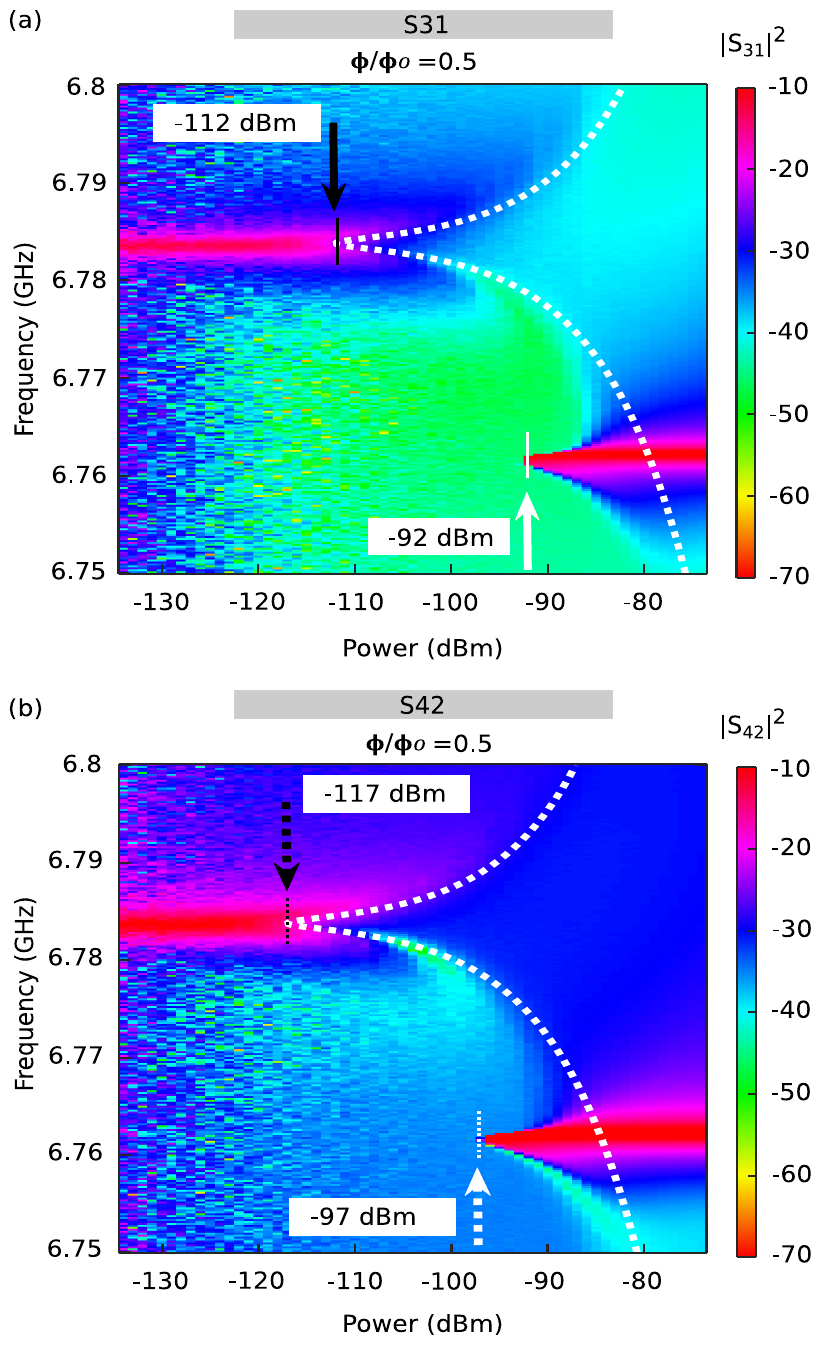}
\caption{Measured transmission coefficient of the device as a function of the injected microwave power and frequency.
(a) Transmission coefficient $|S_{31}|^2$. (b) Transmission coefficient $|S_{42}|^2$. The dotted lines are the theoretically expected positions
of the peak maxima given by Eq. (\ref{Omega_r}).}
\label{Freq_Power}
\end{figure}

\section{Results and discussion}
\label{Experimental observations}

In this section we report on one tone-spectroscopy measurements where we study the qubit-resonator interaction under small applied magnetic fields, and at various injected microwave powers. Transmission rectification manifests itself as the difference between the transmission coefficients $S_{31}$ and $S_{42}$. 
The transmission coefficient $S_{XY}$ 
is defined as the ratio of the signal amplitude coming out of the port X and of that going into the port Y. 
We tune the qubit to the degeneracy point, corresponding to half a flux quanta threading the qubit loop, $\Phi=0.5\Phi_0$.  

We sweep the probe signal frequency at different probing powers. 
To minimise possible errors caused by the attenuation in the input lines, we have applied an in-situ calibration method described in Section ~\ref{Appendix_A} under background calibration subsection.
In Figs. ~\ref{Freq_Power} (a) and ~\ref{Freq_Power} (b)
we plot the transmission coefficient $|S_{31}|^2$ and $|S_{42}|^2$ obtained in this way.
 
We explore the range of frequencies around
the frequency of the hybrid mode of the two resonators $f_h=6.761$ GHz. This mode is formed because the
resonators 1 and 2 are coupled not only to the qubit, but also to each other via the inductances $L_1,L_2$.  
At small microwave powers the frequency of this mode is shifted upwards dispersively induced by its coupling to the qubit, where $\chi=22$ MHz, and
the transmissions $S_{31}$ and $S_{42}$ exhibit resonances at the frequency $f_h+\chi=6.784$ GHz.
At high powers the hybrid mode decouples from the qubit and the resonance frequency moves from $f_h+\chi$ to $f_h$. 
At intermediate power levels the system shows strongly non-linear behavior typical for the quantum Duffing oscillator \cite{Yamamoto, Yamaji, Drummond}.
The lowest power, required to drive the system in the strongly non-linear regime, is observed at resonance frequency $f_h+\chi$.
We denote such power as $P^{\rm nl}_j$, where the index $j$ indicates the resonator through which the driving power is injected.    
In Fig.~\ref{Freq_Power} (a) we indicate the power $P^{\rm nl}_1=-112$~dBm by a solid black arrow, and 
in Fig.~\ref{Freq_Power} (b) we mark $P^{\rm nl}_2=-117$~dBm by a dotted black arrow. The observed power difference equals $5$ dB, or $4.3$ fW. 
It provides power scale of transmission rectification in our device.

\begin{figure*}[ht]
\includegraphics [width = 0.9\textwidth] {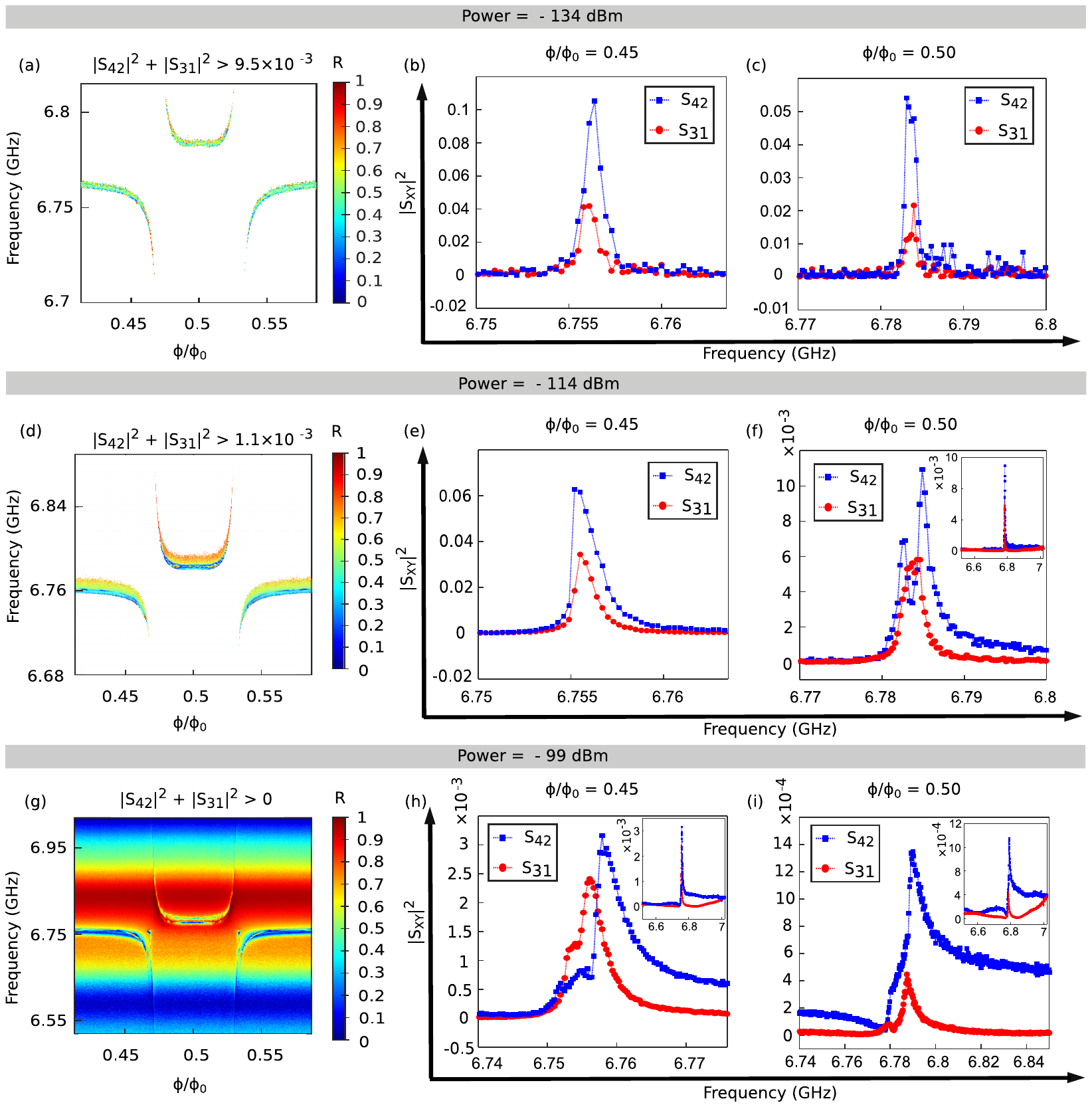} 
\caption{(a,d,g) The transmission rectification ratio R of Eq.(\ref{R}) measured at three different levels of the injected microwave power P, -134 dBm, -114 dBm and -99 dBm. Here, $\Phi_{\rm}$ is the external magnetic flux and $\Phi_0$ $(= h / 2e)$ is the magnetic flux quantum.
(b,e,h) Transmission coefficient $|S_{31}|^2$ and $|S_{42}|^2$  for the same three levels of power and at the flux value $\Phi/\Phi_0=0.45$.
(c,f,i) Transmission coefficient $|S_{31}|^2$ and $|S_{42}|^2$  at the flux value $\Phi/\Phi_0=0.5$. }
\label{Rectification_plot}
\end{figure*}

We define the transmission rectification ratio (R) \cite {F. Fratini, A.V.Marcos}  as
\begin{equation}
\begin{split}
R =  \left|\frac{{|S_{42}|^2}-{|S_{31}|^2}}{{|S_{42}|^2}+{|S_{31}|^2}}\right|
\label{R}
\end{split}
\end{equation}
In Figs. ~\ref{Rectification_plot} (a,d,g) we plot this ratio as a function of frequency and magnetic flux for three different levels of the injected microwave power to show the increasing trend in transmission rectification ratio with the injected power. To suppress background noise, in these figures we retain only such data points
where the sum $|S_{42}|^2+|S_{31}|^2$ exceeds certain threshold value provided in the label above the graphs.
We observe stronger transmission rectification close to the resonance frequencies
corresponding to the hybrid modes of the resonators and the qubit. These modes are also revealed by the usual one tone-spectroscopy measurements,
see Fig. \ref{Supplement_1tone_2tone_fits} in Appendix C. To illustrate the transmission rectification effect further, in  
Figs.~\ref{Rectification_plot} (b,c,e,f,h,i) we plot the transmission coefficient
$|S_{42}|^2$ and $|S_{31}|^2$  at the same three levels of the microwave power and at two selected values of the flux, $\Phi/\Phi_0=0.45$ and $\Phi/\Phi_0=0.5$.
At the lowest power, $-134$ dBm, the transmission rectification happens only very close to the resonance frequencies, see Fig.~\ref{Rectification_plot} (a,b,c). 
In contrast, at high power -99 dBm (see Fig.~\ref{Rectification_plot} (g,h,i)) the transmission rectification ratio exceeds $60\%$ in the wide band of $\approx250$ MHz  near the avoided crossing points, and between $6.81$ GHz to $6.86$ GHz, it exceeds over $90\%$ for a 50 MHz wide range. 
At the intermediate power -114 dBm, we observe splitting of the single line in the transmission coefficient $S_{42}$ into two, whereas the transmission coefficient $S_{31}$ still maintains one line as it does at low microwave powers (see Fig.~\ref{Rectification_plot} (f)).

The origin of the diode effect in our device is the non-linearity of the qubit.
For this reason, single Lorentzian lines in the transmission coefficients $|S_{31}|^2$ and $|S_{42}|^2$ centered at
frequency $f_h+\chi$ split into two lines above the threshold values of the input powers $P_1^*$ and $P_2^*$. 
In our sample $P_1^*>P_2^*$ (see Fig. \ref{Freq_Power}).
The positions of the two peaks after the splitting, i.e. for $P>P_j^*$, are given by the expression
\begin{eqnarray}
f_\pm(P_j) = f_h+\chi \pm \frac{\sqrt{2}}{3}\frac{\kappa_h}{2\pi}\sqrt{\frac{P_j^{\rm in}}{P_j^*}-1}.
\label{Omega_r}
\end{eqnarray}
Here $\kappa_h/(2\pi)=1.1$ MHz is the line-width of the peak at low power. The derivation of Eq. (\ref{Omega_r})
is provided in Appendix \ref{Theory}. In Fig. \ref{Freq_Power} we indicate the peak positions (\ref{Omega_r}) by white dashed lines.
They agree well with the experimentally observed ones.
In the experiment we observe the ratio
$P_1^*/P_2^* =3.2$, which corresponds to a 5 dB difference in power on the log-scale.
A strong transmission rectification effect with $R\sim 1$ occurs at powers $P\gtrsim P_2^*/3$, at which the peaks in $|S_{31}|^2$ and $|S_{42}|^2$
overlap weakly. Theory model of Appendix \ref{Theory} predicts that the ratio $P_1^*/P_2^*$ scales as $P_1^*/P_2^*\propto \kappa_{h2}/\kappa_{h1}$, where
$\kappa_{h1}$ and $\kappa_{h2}$ are the partial contributions to the damping rate of the hybrid mode due to the leakage of the
energy via the capacitors $C_{K1}$ and $C_{K2}$. Therefore, one can further enhance the diode effect in our system by making these
capacitors unequal.

\section{Conclusion}

We propose a flux tunable on-chip microwave diode architecture. 
It is based on a superconducting flux qubit inductively coupled to two superconducting resonators. Using one tone-spectroscopy, tuning the qubit to the degeneracy point by applying the half-flux quantum to it, and performing two separate measurements with microwave signals coming either through resonator 1 or resonator 2, we obtain $5$ dB difference between the powers needed to drive the qubit to the strongly non-linear regime. Furthermore, near the qubit-resonator avoided crossing region we observe high transmission rectification ratio $R>90\%$ for a narrow frequency bandwidth of $50$ MHz, and $R>60\%$ for a wider bandwidth of $250$ MHz, at $-99$ dBm input microwave power. The transmission rectification ratio is flux tunable and strongly depends on the input microwave power.
 Based on the reported resonator-qubit-resonator geometry, the future goal is to realize tunable photonic quantum heat valve  \cite{Ronzani2019, LuY} and  quantum heat rectifier \cite{Robnagel, Jorden2020}, in the field of circuit quantum electrodynamics. Furthermore, the reported strong non-reciprocal behaviour remains relevant for potential applications in quantum information \cite{Ronzani2019, Kosloff, Sai}, microwave read-out components \cite{A.R. Hamann, Archana Kamal, Lecocq} and optomechanics \cite{Juha, Barzanjeh}.

\section{Methods}

\label{Appendix_A}

\subsection *{Fabrication}
\label{Fabrication}

The device was fabricated on a highly resistive 675~$\mu$m thick Si wafer covered by 30~nm thick dielectric layer of Al$_2$O$_3$, which has been made using atomic layer deposition (ALD). In the next step, a 200~nm thick layer of superconducting niobium (Nb) has been sputtered using DC magnetron sputtering technique. This step has been followed by patterning of the feed lines, resonators and ground-planes on a 300~nm thick positive electron beam resist. Post baking is done at 150~$^{\circ}$C for 5 minutes, followed by the reactive ion etching (RIE) using CF$_4$ $+$ O$_2$ chemistry. 
Post baking step  allows the resist to improve the adhesion with the substrate thus minimizing the chances of etching of unexposed Nb parts. 
In the second lithography step, a bi-layer PMMA/MMA resist has been used to write Josephson junctions using the standard Dolan bridge technique \cite{Dolan}. 
The exposed substrate has been developed in Methyl-Isobutyl-Ketone (MIBK):
Isopropanol alcohol (IPA) developer solution, and Methyl-glycol-Methanol solution, respectively. 
Metal deposition has been performed using an e-beam evaporator. 
In-situ argon plasma milling has been performed on the sample surface to mill the native oxide before the evaporation.
It has been done to obtain a clean contact between Nb layers and Al layers. 
Afterwards, a 30~nm thick aluminium metal layer has been evaporated at $+18^{\circ}$ and it has been oxidized
in-situ to create the tunnel barrier in the junctions. 
Subsequently, a second 30~nm thick aluminium layer is evaporated at $-18^{\circ}$. To strip the deposited metal from the unexposed areas, samples are immersed in hot acetone for 40 minutes. The sample is visually examined under electron microscopy before proceeding to dicing and measurements. 
The room temperature resistance of the test SQUID fabricated on the same chip to mimic the real device has been measured to be $R \approx 2.6 ~\mathrm{k\Omega}$.

\begin{figure*}[ht]
\includegraphics [width = 0.9\textwidth] {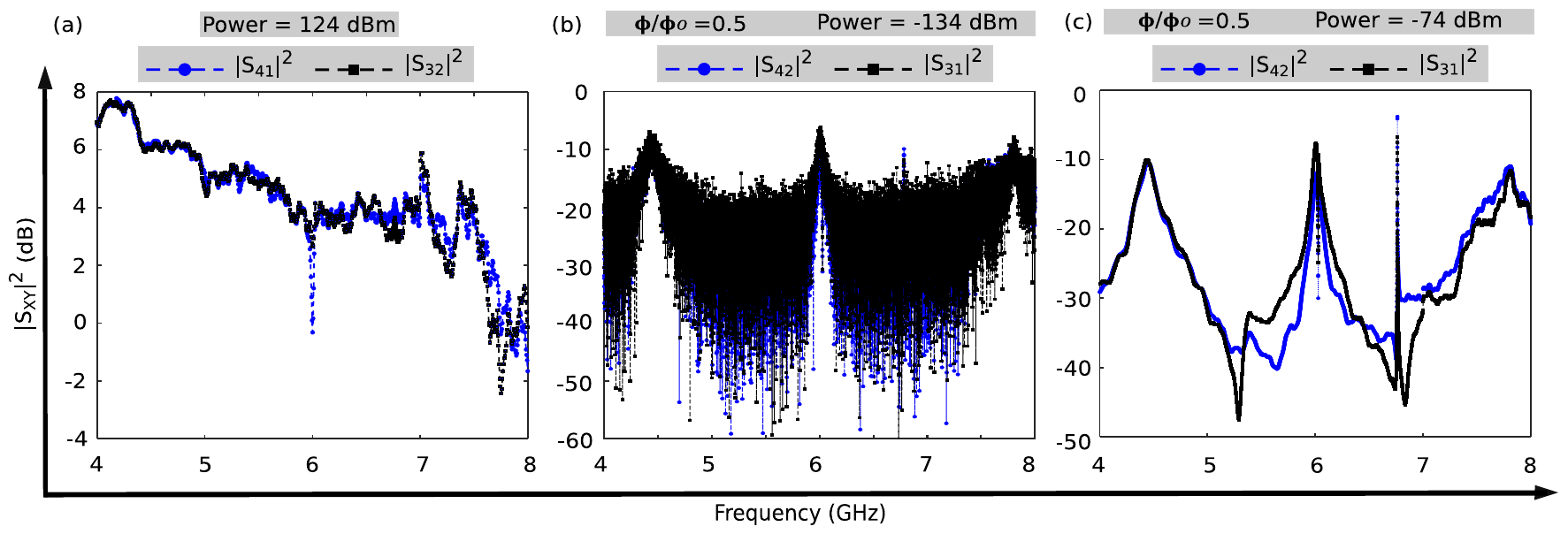} 
\caption{Wide-band transmission coefficient graphs. (a). Transmission coefficient plots for sides S41 and S32, reconstructed to obtain transmission baseline for background calibration. (b,c). Calibrated transmission coefficient plots obtained from Eq. (\ref{S31}) and Eq. (\ref{S42}), measured across the device at -134 dBm and -74 dBm microwave power.}
\label{Supp_Calib}
\end{figure*}

\subsection *{Background calibration}
\label{Background calibration}

To ensure the accuracy of the measurements, we perform an in-situ calibration as described below. Let us consider the measured transmission coefficient $S_{O_{4}I_{1}}$ from input 1 ($I_1$) to output 4 $(O_{4})$ (see Fig. \ref{Measurement_setup} (b)). 
It can can be expressed as
\begin{equation}
\begin{split}
    \tilde S_{O_{4}I_{1}} = \tilde S_{1I_{1}} + \tilde S_{41} + \tilde  S_{O_{4}4}.
\label{SO4I1}
\end{split}
\end{equation}
Here $\tilde S_{ij}=10\log_{10}(|S_{ij}|^2)$ is the transmission coefficient in decibel (dB), 
$\tilde S_{1I_{1}}$ is the attenuation factor from the input line 1 to the port 1 of the device,
$\tilde S_{41}$ is the transmission coefficient between the ports 1 and 4, which we are looking for, and $\tilde S_{O_{4}4}$ is
the attenuation between the port 4 of the device and the input port of the microwave switch.

Similarly, we can write the transmission from the input 2 ($I_2$) to the output 3 $(O_{3})$ as
\begin{equation}
\begin{split}
    \tilde S_{O_{3}I_{2}} =  \tilde S_{2I_{2}}+ \tilde S_{32} +  \tilde S_{{O_{3}3}},
\label{SO3I2}
\end{split}
\end{equation}
and the transmissions across the device as
\begin{equation}
\begin{split}
    \tilde S_{O_{3}I_{1}} = \tilde S_{1I_{1}} + \tilde S_{31} +  \tilde S_{O_{3}3},
\label{SO3I1}
\end{split}
\end{equation}
\begin{equation}
\begin{split}
    \tilde S_{O_{4}I_{2}} = \tilde S_{2I_{2}} + \tilde S_{42} +  \tilde S_{O_{4}4}.
\label{SO4I2}
\end{split}
\end{equation}

We connect the ports 3 and 4 of the device to the input ports $O_{3}$ and $O_{4}$ of the microwave switch with the identical relatively short wires. 
Therefore, we could assume that $\tilde S_{O_{3}3}$ $=$ $\tilde S_{O_{4}4} =0$.

Next, we perform spectroscopy at off-resonance frequencies at different flux points. Due to the strong qubit-resonator coupling the change in the dispersive shift under the applied magnetic flux is larger than the line-width of the resonance. 
Therefore, by sweeping the magnetic flux, we can tune the device to the off-resonance regime with $\tilde S_{41}= \tilde S_{32} = 0$ 
for each frequency and in this way can
measure the background transmission coefficient 
$\tilde S_{O_4 I_1 {\rm bg}} = \tilde S_{1I_{1}}  +  \tilde S_{O_{4}4}$
and $\tilde S_{O_3 I_2 {\rm bg}}=\tilde S_{2I_{2}} +  \tilde S_{O_{3}3}$ 
for the entire range of frequencies relevant for the experiment.
They are plotted in Fig. \ref{Supp_Calib} (a).
Subtracting Eq. (\ref{SO4I1}) from Eq. (\ref{SO3I1}), Eq. (\ref{SO3I2}) from Eq. (\ref{SO4I2}), and recalling that $\tilde S_{O_{3}3}$ $=$ $\tilde S_{O_{4}4} =0$,
we obtain the calibrated transmission coefficients of our system as
\begin{equation}
    \tilde S_{31} = \tilde S_{O_{3}I_{1}} 
    - \tilde S_{O_4 I_1 {\rm bg}},
\label{S31}
\end{equation}

\begin{equation}
    \tilde S_{42}  = \tilde S_{O_{4}I_{2}} 
    -  \tilde S_{O_3 I_2 {\rm bg}}.
\label{S42}
\end{equation}
The same method has been used to calibrate the transmissions $S_{41},S_{32}$. 

We also note, that even without calibration
the maximum observed difference between the background transmissions 
is less than $10\%$, while the observed difference in characteristic input powers $P_1^*$ and $P_2^*$ is more than $50 \%$. 
The calibrated transmission coefficients obtained from Eq. (\ref{S31}) and Eq. (\ref{S42}) at two different microwave powers are plotted in Fig.~\ref{Supp_Calib} (b), (c). In Fig.~\ref{Supp_Calib} (c), corresponding to the high microwave power ($-74$ dBm) applied to the device at the magnetic flux $\Phi=0.5\Phi_0$, 
we observe that $\tilde S_{31}$ and $\tilde S_{42}$ almost coincide in the frequency range from $4$ GHz to $5.2$ GHz, the difference between them 
does not exceed 1.5 dB. 
In contrast, in the range from $5.2$ GHz to $7.8$ GHz we observe significant difference, i.e. the diode effect, which reaches 15 dB near the frequency of the hybrid mode.

\section{Acknowledgment} 

We acknowledge the financial support from Academy of Finland grants (grant number 297240, 312057 and 303677), and from the European Union’s Horizon 2020 research and innovation programme under the European Research Council (ERC) programme (grant number 742559) and Marie Sklodowska-Curie actions (grant agreements 766025).
We sincerely recognize the provision of facilities by Micronova Nanofabrication Centre, and OtaNano - Low Temperature Laboratory of Aalto University which is a part of European Microkelvin Platform EMP (grant number. 824109), to perform this research. We thank and acknowledge VTT Technical Research Center for provision of high quality sputtered Nb films.

\section{Author contributions}

The conceptual idea was drafted by R.U. and G.T. The device design and fabrication was done by R.U. The measurements were carried by R.U. The data analysis is conducted by R.U., Y-C.C. and D.S.G. The theoretical model was conceived by D.S.G., in collaboration with G.T. The microwave simulations were carried by Y-C.C. and A.G. In this work D.S.G. and Y-C.C. contributed equally. The technical support was provided by J.T.P. The work was supported and supervised by J.P.P. The manuscript was written by R.U., with important contributions from all the authors.

\appendix

\section {Simulations}
\label{Simulations}

\begin{figure*}[ht]
\includegraphics [width = 0.7\textwidth] {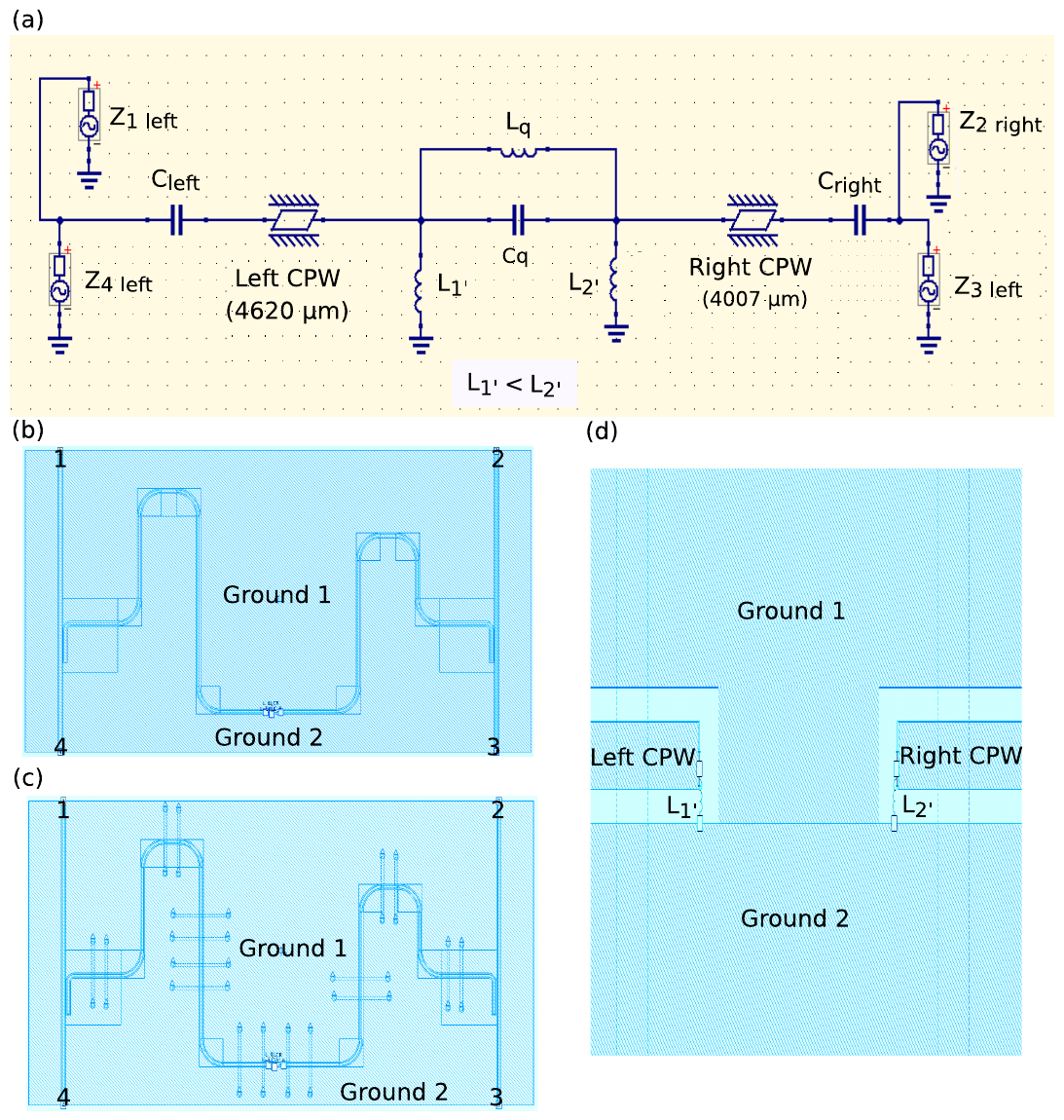}
\caption{(a) Simulated circuit design image from QUCS, multiports network analysis. (b) Simulated design image using Sonnet with no bonding wires. (c) Al Bonding wires mimicking the real device in panel to connect upper (ground 1) and lower (ground 2) grounds. (d). A metallic block to connect the upper and lower ground planes.
}
\label{Simulation_designs}
\end{figure*}

We used lumped elements circuit simulator (QUCS), and high-frequency RF/MW electromagnetic analysis (SONNET), to simulate the two meandering wire shunted resonators to ensure the resonances we measured are reasonable, as shown in Fig.~\ref{Simulation_designs}. The lengths of the CPW waveguide part of the resonators are 4620 um on the left and 4007 um on the right. The corresponding inductance of the left ($L_{1'}$) and right ($L_{2'}$) meandering wire is $\approx$ 0.1 nH and $\approx$ 0.25 nH \cite{R.Upadhyay}.  From finite element analysis simulator (COMSOL Multiphysics), we calculate the coupling capacitance value of 6.98 fF between the transmission line and the resonator ($C_{left}$ and $C_{right}$), which we use in simulations. The simulated resonance frequencies from QUCS were 6.072 GHz and 6.394 GHz. Simulations from Sonnet reveals the frequencies of the center lines at 5.311 GHz and 6.130 GHz. Since in the reported device we use air bridges to connect the ground planes around the circuit, therefore we now repeat the simulation in Sonnet using the air bridges. In superconducting quantum devices, air bridges \cite{Abuwasib, Janzen} holds the balance of the ground planes around the central lines, reduces the possibility of microwave loss due to mode mixing, and avoids pseudo resonances.  The simulated designs with and without air bridges are reported in Fig.~\ref{Simulation_designs} (b,c).  After adding the air bridges to mimic the reported device, the simulated frequencies rise to 5.706 GHz and 6.182 GHz. Both of these frequencies are lower than the designed bare frequencies of the left and right resonators, this could be due to the imperfect isolation of the propagating microwave. We observed that the simulated current-field distribution at the resonance frequency at one side of the device shows a part of the field leaking across the other side because of the imbalance of the grounding around the meandering shunt that could make the wave to propagate along the ground surrounded by the centre line of the off-resonance resonator. This imbalance issue could be verified when we placed a perfect conductive block as shown in Fig.~\ref{Simulation_designs} (d), between the meandering wire to connect the upper (ground 1) and lower (ground 2) ground planes of the device, and as a result we observed that the frequencies rise to the values very close to QUSC. The experimentally observed lower frequency (6.027 GHz) is within a reasonable range obtained from Sonnet and QUCS simulations, but the higher frequency (6.762 GHz) is higher than the simulated frequency from QUCS. As shown in Fig.~\ref{Simulation_designs} (a), adding an inductor with around 0.5 nH between the floating sides of the meandering wires to mimic the Josephson junction could increase the frequency to 6.7 GHz. Hence, both of our measured frequencies are under the reasonable range and are used in Appendix \ref{Theory} for calculations.

\section{Theory}
\label{Theory}

\begin{figure*}[ht]
\includegraphics [width = 0.9\textwidth] {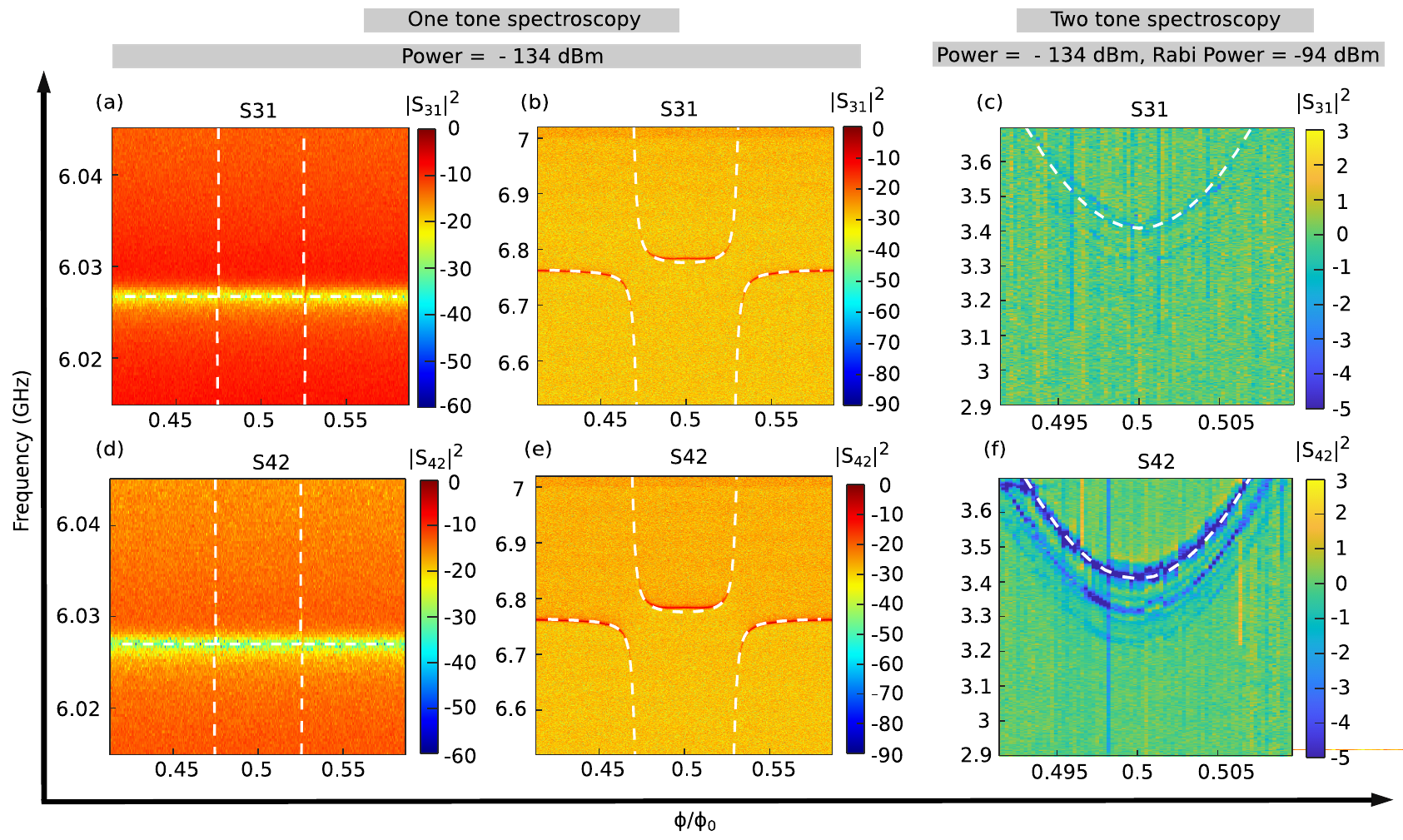} 
\caption{One tone and two tone-spectroscopy for -134 dBm input probing power. 
Panels (a) and (b) show the one tone-spectroscopy data, and panel (c) --- two tone-spectroscopy data, 
with probe signal coming through the port 1. 
Analogously, panels (d) and (e) show one tone-spectroscopy  and (f) --- two tone-spectroscopy, with the probe signal arriving through the port 2. 
In panels (c) and (f) the pump tone power (Rabi power) is -94 dBm.}
\label{Supplement_1tone_2tone_fits}
\end{figure*}

In this Appendix we briefly present the theory of the transmission rectification effect in our device.
The system is described by the Hamiltonian
\begin{eqnarray}
H=H_{\rm res} + H_{\rm loop} + H_{\rm int}.
\label{H_system}
\end{eqnarray}
The Hamiltonian of the two resonators is given by
\begin{eqnarray}
H_{\rm res} &=& \sum_{j=1}^2\left[ \hbar \omega_ja_j^\dagger a_j
+ \sqrt{\frac{\hbar\kappa_{cj}Z_0}{\omega_j^3}}\frac{dI_j(t)}{dt}(a_j^\dagger + a_j)\right]
\nonumber\\&&
-\,\hbar g_{12}(a_1^\dagger+a_1)(a_2^\dagger+a_2),
\label{Hresonator}
\end{eqnarray}
where $\omega_j=2\pi f_j$ are the angular frequencies of the fundamental modes of the resonators 1 and 2 (in our sample $\omega_2>\omega_1$), 
$a_j$ are the corresponding ladder operators, 
$\kappa_{c1},\kappa_{c2}$ are the damping rates of the resonators due to their capacitive coupling to the transmission lines, 
\begin{eqnarray}
\kappa_{cj}=\frac{2\omega_j^3Z_0^2C_{Kj}^2}{\pi},
\label{kappa_c}
\end{eqnarray}
$Z_0$ is the resonator impedance, $C_{K1},C_{K2}$ are the capacitors connecting the resonators to the transmission lines 
(see Fig. \ref{Device_concept}c),
$I_1(t),I_2(t)$ are the input microwave currents which are related to the incoming powers as $P_j^{\rm in}=Z_0\overline{I_j^2(t)}$ 
(here bar implies the time averaging), 
and $g_{12}$ is the coupling strength between the two resonators.
The Hamiltonian of the superconducting loop $H_{\rm loop}$ is expressed in terms of the ladder operators 
$b, b^\dagger$ describing its low-frequency mode, and the flux dependent qubit frequency $\omega_0(\Phi)=2\pi f_{01}(\Phi)$, 
\begin{eqnarray}
H_{\rm loop} = \hbar \omega_0(\Phi) b^\dagger b - \frac{E_C}{12}(b^\dagger+b)^4.
\label{Hloop}
\end{eqnarray}
Here $E_C$ is the effective charging energy of the low frequency mode of the loop, which determines the anharmonicity of the qubit.
From the spectroscopic measurements presented in Fig. \ref{Supplement_1tone_2tone_fits} (e,f) we estimate $E_C/(2\pi\hbar)\approx 200$ MHz.
Finally, the last term in the Hamiltonian (\ref{Hinteraction}) describing the interaction between the loop and the resonators  has the form 
\begin{equation}
H_{\rm int} = -\hbar g_1(a_1^\dagger+a_1)(b^\dagger+b) -\hbar g_{2} (a_2^\dagger+a_2)(b^\dagger+b),
\label{Hinteraction}
\end{equation}
where $g_1$ and $g_2$ describe the coupling between the qubit and the corresponding resonator.

We diagonalize the Hamiltonian of the two resonators (\ref{Hresonator}) and introduce the hybrid modes
with frequencies 
\begin{eqnarray}
\omega_{h,l} = \sqrt{\frac{\omega_1^2+\omega_2^2\pm\sqrt{(\omega_2^2-\omega_1^2)^2 + 16 g_{12}^2\omega_1\omega_2}}{2}}.
\label{omega_hl}
\end{eqnarray}
In the transmission rectification experiment we probe the range of frequencies close to $f_h=\omega_h/(2\pi)=6.762$ GHz. For this reason,
we leave only the high frequency mode and write the Hamiltonian of the resonators (\ref{Hresonator}) and 
the interaction term (\ref{Hinteraction}) in the form
\begin{eqnarray}
H_{\rm res} &=& \hbar\omega_h a_h^\dagger a_h + 
\left( \frac{\sqrt{\kappa_{h1}}}{\omega_1^2}  \frac{dI_1(t)}{dt}
\right.
\nonumber\\ &&
\left.
+\, \frac{\sqrt{\kappa_{h2}}}{\omega_2^2}  \frac{dI_2(t)}{dt} \right)\sqrt{\hbar\omega_h Z_0} (a_h^\dagger + a_h),
\label{Hres}
\\ 
H_{\rm int} &=& - \hbar g_h  (a_h^\dagger + a_h)(b^\dagger + b).
\label{Hint}
\end{eqnarray}
Here $a_h,a_h^\dagger$ are the ladder operators of the hybrid mode with the angular frequency $\omega_h=2\pi f_h$, 
$\kappa_{h1}$ and $\kappa_{h2}$ are the partial contributions to the total damping rate of this mode
coming from the coupling of the resonators 1 and 2 to the transmission lines,
\begin{eqnarray}
\kappa_{h1} = \frac{\omega_1}{\omega_h}\sin^2\theta\,\kappa_{c1}, \;\;\; 
\kappa_{h2} = \frac{\omega_2}{\omega_h}\cos^2\theta\,\kappa_{c2},
\label{kappa_h}
\end{eqnarray}
and the angle $\theta$ is determined by
\begin{eqnarray}
\sin 2\theta = \frac{4g_{12}\sqrt{\omega_1\omega_2}}{\sqrt{(\omega_2^2-\omega_1^2)^2 + 16 g_{12}^2\omega_1\omega_2}}.
\label{theta}
\end{eqnarray}
The total damping rate of the hybrid mode is $\kappa_h=\kappa_{h1}+\kappa_{h2}+\kappa_{hi}$, where $\kappa_{hi}$
describes the internal damping in the resonators.

Based on (\ref{Hres}) we observe that the input currents $I_1^{\rm in}$ and $I_2^{\rm in}$ have different
pre-factors. This difference is the origin of the asymmetry in our system. From this pre-factors we can determine the ratio
of the powers $P_1^*$ and $P_2^*$, above which single transmission lines in the coefficients $|S_{31}|^2$ and $|S_{42}|^2$ split into two lines, 
without solving the problem. Namely, we find 
\begin{eqnarray}
\frac{P_1^*}{P_2^*} = \frac{\omega_1^4\kappa_{h2}}{\omega_2^4\kappa_{h1}}.
\label{ratio}
\end{eqnarray}
The powers $P_1^*$ and $P_2^*$ are not equal because the frequencies of the two resonators $f_1$ and $f_2$ and/or
the couplings between the hybrid mode and the two transmission lines, $\kappa_{h1}$ and $\kappa_{h2}$, differ from each other.
This makes our device a non-linear system with asymmetric coupling to the two ports. According to the theory, exactly these properties
are required for transmission rectification, see e.g. Ref. \cite{Segal}. In our device the coupling capacitors $C_{K1}$ and $C_{K2}$ are nominally equal, $C_{K1}=C_{K2}$.
Therefore the ratio (\ref{ratio}) can be simplified to $P_1^*/P_2^*=\cot^2\theta$. In the experiment we find $P_1^*/P_2^*=3.2$, 
see Fig. \ref{Freq_Power}, which corresponds to $\theta=0.51$. To obtain this value from Eq. (\ref{theta}) and, at the same time, to reproduce
the experimentally observed frequencies of the modes $f_h=6.762$ GHz and $f_l=6.026$ GHz from Eq. (\ref{omega_hl}), we choose
$f_1=6.209$ GHz, $f_2=6.595$ GHz and $g_{12}/(2\pi)=313$ MHz. The simulations reported in Appendix \ref{Simulations} show that these parameters are reasonable.
To verify our model further, we have estimated the values of the coupling capacitors in COMSOL and found $C_{K1}=C_{K2}=7$ fF.
Adopting this value and the parameters given above, from
Eq. (\ref{kappa_c}) we obtain $\kappa_{c1}/(2\pi)=732.6$ kHz, $\kappa_{c2}/(2\pi)=878$ kHz, and from Eq. (\ref{kappa_h}) we find
$\kappa_{h1}/(2\pi)=160$ kHz, $\kappa_{h2}/(2\pi)=653$ kHz. 
The damping rates $\kappa_{h1},\kappa_{h2}$ can be independently estimated by fitting the transmission coefficients $|S_{41}|^2$ and $|S_{32}|^2$
at zero magnetic flux, where the qubit is decoupled from the resonator, to the expressions resulting from our model
Hamiltonian (\ref{Hresonator}):
\begin{eqnarray}
|S_{41}|^2 &=& 1 - \frac{\omega_h^4}{\omega_1^4}\frac{ \kappa_{h1}^2  + 2\kappa_{h1}\kappa_{h2}}{4(\omega-\omega_h)^2+\kappa_h^2},
\nonumber\\
|S_{32}|^2 &=& 1 - \frac{\omega_h^4}{\omega_2^4}\frac{ \kappa_{h2}^2 + 2\kappa_{h1}\kappa_{h2}}{4(\omega-\omega_h)^2+\kappa_h^2}.
\label{S41_th}
\end{eqnarray}  
Such fitting procedure gives $\kappa_{h1}/(2\pi)=160$ kHz, $\kappa_{h2}/(2\pi)=430$ kHz and $\kappa_h/(2\pi)=787$ kHz.
While $\kappa_{h1}$ agrees with the theoretical estimate given above, the experimental rate $\kappa_{h2}$ is a bit lower than the theoretical prediction.
With these parameters the ratio of the powers (\ref{ratio}) becomes $2.1$, which is still not very far from the result of the measurements. 
Thus, we have confirmed that our observations reasonably well agree with the model.

The dependence of the transmission coeffcients $|S_{31}|^2$ and $|S_{42}|^2$ on power shown in Fig. \ref{Freq_Power} can be understood
as follows. At sufficiently low power we can approximately replace the two resonators and the qubit by a single non-linear system
with the Hamiltonian
\begin{eqnarray}
H &=& \hbar \omega_r a^+a + \frac{\hbar K}{6} (a^\dagger + a)^4 
\nonumber\\ &&
+\, 2\hbar(\epsilon_1+\epsilon_2)\cos\omega t(a^\dagger +a).
\label{H_app}
\end{eqnarray}
Here $\omega_r = \omega_h+2\pi\chi$ is the frequency of the hybrid mode shifted due to the interaction with the qubit,
$K$ is the Kerr non-linearity of the combined system and
\begin{eqnarray}
\epsilon_j = \frac{\omega_h^2}{\omega_j^2}\sqrt{\frac{\kappa_{hj} P_j^{\rm in}}{2\hbar\omega_h}}.
\end{eqnarray}
It has been experimentally shown that this approximation well describes systems similar to ours \cite{Yamamoto,Yamaji}.
To find the the transmission coefficient $|S_{31}|^2$ at low power we put $\epsilon_2=0$ and constract the following ratio:
\begin{eqnarray}
|S_{31}|^2 = \frac{P_3}{P_1^{\rm in}} = \frac{\kappa_{h2}\hbar\omega_r}{2P_1^{\rm in}}\langle a^\dagger a\rangle.
\end{eqnarray}
Here $P_3$ is the power coming to the port 3 and 
the factor 2 in the denominator accounts for the equal sptitting of the power coming out of the resonator 2  
between the ports 2 and 3, i.e. we put $\kappa_{h2}\hbar\omega_r\langle a^\dagger a\rangle = P_2+P_3$ and assume that $P_2=P_3$. 
The average value $\langle a^\dagger a\rangle$ for the Hamiltonian (\ref{H_app}) has been evaluted in Ref. \cite{Drummond}. 
Based on this result we obtain the expression for the transmission coefficient in the form
\begin{eqnarray}
&& |S_{31}|^2 \approx  \frac{\kappa_{h1} \kappa_{h2} }{4(\omega-\omega_r)^2+\kappa_h^2}
\nonumber\\ && \times\,
\frac{\left|_0F_2\left(1-\frac{\omega-\omega_r-i\frac{\kappa_h}{2}}{K},-\frac{\omega-\omega_r+i\frac{\kappa_h}{2}}{K},\frac{2\kappa_{h}^2 P_1^{\rm in}}{9 K^2 P_1^*}\right)
\right|^2}
{\left|_0F_2\left(-\frac{\omega-\omega_r-i\frac{\kappa_h}{2}}{K},-\frac{\omega-\omega_r+i\frac{\kappa_h}{2}}{K},\frac{2\kappa_{h}^2 P_1^{\rm in}}{9 K^2 P_1^*}\right)\right|^2}, \;\;\;\;\;\;\;
\label{S31_th}
\end{eqnarray}
where $_0F_2(x)$ is the generalized hypergeometric function and
\begin{eqnarray}
P_1^* = \frac{2}{9}\frac{\hbar\kappa_h^2\omega_1^4}{\kappa_{h1}\omega_h^3}
\label{P1*}
\end{eqnarray}
is the power at which the Lorentzian peak in $|S_{31}|^2$ splits into two. The transmission coefficient $|S_{42}|^2$ and the power $P_2^*$
are given by the same experssions with the interchanged indexes 1 and 2. Analyzing the expression (\ref{S31_th}),
one can show that in the limit $\kappa_h\ll K$ the maxima of the
two peaks appearing at $P_1^{\rm in}>P_1^*$ occur at frequencies  
\begin{eqnarray}
\omega_\pm = \omega_r \pm \frac{\sqrt{2}\kappa_h}{3}\sqrt{\frac{P_1^{\rm in}}{P_1^*}-1}.
\end{eqnarray}
Inverting this formula, we find that for a given probe frequency $\omega$ the peak in the transmisssion coefficient occurs at the power
\begin{eqnarray}
P_{\rm peak,1} = P_1^*\left(1+\frac{9(\omega-\omega_r)^2}{2\kappa_h^2}\right).
\label{P_peak}
\end{eqnarray}

In Fig. \ref{Freq_Power} we show the powers $P_{\rm peak,1}$ and $P_{\rm peak,2}$ by the white dashed lines.
We have used the damping rate $\kappa_h=1.1$ MHz, which has been obtained by fitting the transmission coefficients (\ref{S41_th})
at low powers and at the flux value $\Phi=0.5\Phi_0$, at which the data shown in Fig. \ref{Freq_Power} have been gathered.
The threshold powers $P_1^*=-112$ dBm and $P_2^*=-117$ dBm have been treated as fitting parameters.
From the fits we also estimate the anharmonicity as $K/(2\pi)\approx -11.5$ MHz.

Finally, we note that at sufficently high power the assumption about weak non-linearity of the qubit becomes insufficient,
and one should consider full sinusoidal current-phase relation for the three Joesephon junctions of the qubit.
Here we do not consider this regime. It is well known, however, that in this limit the resonator becomes decoupled from the qubit,
and the transmission lines both in $|S_{31}|^2$ and in $|S_{42}|^2$ shift to the bare frequency of the hybrid mode $f_h$, see Fig. \ref{Freq_Power}. 

\section{Two tone-Spectroscopy}
\label{Two tone-Spectroscopy}

To determine the qubit transition frequencies we have performed the two-tone spectroscopy as follows. For every value of magnetic flux we choose the probe frequency of a continuous weak microwave signal `probe tone' (tone one) using a vector network analyzer (VNA). Once the flux specific probe frequency is chosen, using a separate microwave signal generator a `pump tone' (tone two) is applied to excite the qubit energy levels. The combined results of the one tone and the two tone spectroscopies are presented in Fig.  \ref{Supplement_1tone_2tone_fits}.
 
To fit the obtained spectra numerically and to find the coupling constants between the qubit and the resonators, we adapted the following procedure. 
First, we diagonalize the Hamiltonian of the flux qubit using two dimensional plane waves as described in article \cite{R.Upadhyay}. 
In this way, we find the dependence of the transition frequency $f_{01}(\Phi)$ between the lowest and the first excited states 
of the qubit decoupled from the resonators on the magnetic flux $\Phi$. Fitting the obtained $f_{01}(\Phi)$ dependence to
the results of spectroscopy away from the anti-crossing points, we estimate the parameters of the flux qubit. Namely, we find the asymmetry parameter $\alpha=0.632$ and the Josephson energies of the two bigger junctions of the SQUID loop,  $E_J/(2\pi\hbar) = I_C/4\pi e =37.5 $ GHz. 
Next, we use the equation describing two coupled oscillators, 
\begin{eqnarray}
f_r = \sqrt{\frac{f_i^2 + f_{01}^2 \pm \sqrt{\left(f_i^2 - f_{01}^2\right)^2 + 16 g_i^2 f_if_{01}}}{2}},
\label{fr}
\end{eqnarray}
to fit every avoided crossing between the hybrid modes of the resonator (indicated by the index $i$)
and the flux dependent qubit frequency in  Fig.  \ref{Supplement_1tone_2tone_fits}.
In the range of frequencies shown there we observe two hybrid resonator modes, low frequency mode (index $l$) and high frequency mode (index $h$).
The frequencies of these modes are $f_l=6.027$ GHz and $f_h=6.762$ GHz, 
and the corresponding couplings are $g_l \approx 1$ MHz and $g_h\approx175$ MHz.

\end{document}